\documentclass[]{aa}
\usepackage[colorlinks=true, linkcolor=blue, citecolor=blue, urlcolor=blue]{hyperref}
\usepackage{amsmath}
\usepackage{graphicx}
\usepackage{amssymb}
\usepackage{mathrsfs}
\usepackage{caption}
\usepackage{float}
\usepackage{afterpage}
\usepackage{amstext}

\usepackage{multirow}

\usepackage{txfonts}
\usepackage{natbib}

\begin{document}

   \title{Stellar Population Astrophysics (SPA) with the TNG}
\subtitle{CNO abundances in 28 open
clusters
}
   \author{
   B. Ćurjurić\inst{1},
   A. Drazdauskas\inst{1},
   G. Tautvai\v{s}ien\.{e}\inst{1},
   A. Bragaglia\inst{2},
   N. Alvarez-Baena\inst{3,4},
   V. D'Orazi\inst{3,4},
      M. Dal Ponte\inst{5}
}

   \institute{Vilnius University, Faculty of Physics, Institute of Theoretical Physics and Astronomy,  Saul\.{e}tekio av. 3, 10257 Vilnius, Lithuania\\
              \email{bruno.curjuric@ff.stud.vu.lt }
    \and
 INAF - Osservatorio di Astrofisica e Scienza dello Spazio, via P. Gobetti 93/3, 40129 Bologna, Italy
   \and
Department of Physics, University of Rome Tor Vergata, via della Ricerca Scientifica 1, 00133 Rome, Italy
   \and
   INAF - Osservatorio Astronomico di Roma, via Frascati 33, Monte Porzio Catone, 00078 Italy
                 \and
   INAF - Ossevatorio Astronomico di Padova, vicolo dell’Osservatorio 5, 35122 Padova, Italy
 }

   \date{Received 30 January 2026 / Accepted 12 April 2026}

\authorrunning{B. Ćurjurić et al.}
\titlerunning {Stellar Population Astrophysics (SPA) with the TNG}

  \abstract
  {}
   {Our main aim with this work was to enlarge the pool of open clusters with determined carbon, nitrogen, and oxygen abundances in evolved giants to further advance chemical clocks in stellar age determinations.  
}
   {High-resolution spectra were analysed using a differential model atmosphere method. Carbon abundances were derived using spectral synthesis of the ${\rm C}_2$ band heads at 5135 and 5635.5~\AA. The CN features at 6470--6490~{\AA} were analysed to determine the abundances of nitrogen. The oxygen abundances were determined from the [O\,{\sc i}] line at 6300~{\AA}. 
} 
{We provide abundances of C, N, and O for 88 giants in 28 open clusters and in two stars of the association Theia\,1214. The results were compared with theoretical predictions and used for the analysis of the [C/N] relations with age, and we investigated the origin of the two Theia\,1214 stars. 
    }
  {Precise age dating requires separate age calibrations of [C/N] ratios for the  
first-ascent giants of the lower part of the red giant branch and for the red clump stars to account for the additional abundance alterations during the post-red giant branch luminosity bump evolution.
In the first-ascent giant stars with larger turn-off masses, the
observed C/N ratios are slightly higher than those predicted by standard models, and  the obtained [C/N] versus age relation is flatter in the young age regime than in previous studies.   
We doubt that the two stars we investigated in the Theia\,1214 association belong to the tail of NGC\,752. Most probably, they are field stars. 
}


   \maketitle

\section{Introduction}

Open clusters are chemically homogeneous groups of stars that form nearly simultaneously from the same molecular cloud \citep[][and references therein]{shu1987, lada2003, Krumholz19}. Consequently, their member stars share a common age, initial chemical composition, and a range of stellar masses. This makes open clusters excellent laboratories for studying stellar evolution, as the differences observed among their stars primarily reflect mass-dependent evolutionary effects rather than variations in age or composition \citep[e.g.][]{friel95, bovy2016, poovelil2020, Lagarde2024}.

Because all cluster members are coeval, fitting theoretical isochrones to a colour–magnitude diagram of an open cluster provides one of the most reliable methods for determining the stellar ages and distances of its members (\citealt{gallart2005, vandenberg13}). The presence of well-defined evolutionary features, such as the main-sequence turn-off, subgiant branch, and red giant branch (RGB), allows for precise constraints on cluster ages, particularly when combined with high-quality photometry and spectroscopic metallicities. Distances can be determined simultaneously through isochrone fitting or independently via parallaxes, making open clusters key benchmarks for calibrating stellar models and distance scales \citep[e.g.][]{bossini19, 2020A&A...640A...1C, dias2021}.

Determining the ages of field stars is considerably more complex \citep[see the review by][]{2010ARA&A..48..581S}. A semi-fundamental method for stellar age estimation is one that relies on a minimal number of assumptions that are both well founded and have limited impact on the derived ages. An example of a semi-fundamental method is nucleocosmochronometry, which measures the abundance of long-lived radioactive isotopes and their decay products (\citealt{cowan1991, 2023ApJ...948..122S}). Some methods are model-dependent, such as isochrone fitting, which compares a star’s position on the Hertzsprung-Russell diagram to theoretical tracks (see e.g. \citealt{2018A&A...618A..54M} for the application to $Gaia$ data and \citealt{sanders2018}), and asteroseismology, which analyses oscillations within a star to infer its internal structure, mass, and hence age (\citealt{chaplin2013, 2020NatAs...4..382C}). Other approaches are empirical, such as gyrochronology, which uses stellar rotation periods as age indicators (\citealt{skumanich1972, barnes2003, 2024AJ....167..159L}), or lithium depletion during the main-sequence phase (e.g. \citealt{herbig1965, jeffries23}. Specific techniques are optimised for certain evolutionary stages, and together these methods provide complementary insights into stellar ageing across diverse stellar populations. 

In this context, chemical abundances have emerged as powerful independent age diagnostics, broadly termed `chemical clocks'. These can be categorised into two distinct physical mechanisms. The first relies on galactic chemical evolution, where ratios of elements produced on different timescales (e.g. [$\alpha$/Fe] or [Y/Mg]) trace the chemical enrichment history of the interstellar medium, thus serving as a proxy for birth time. For example, as the Galaxy ages, the [$\alpha$/Fe] ratio changes \citep[e.g.][] {2012A&A...542A..84D,2013A&A...560A.109H,2019A&A...624A..78D}. Furthermore, this approach has been extended to other elemental ratios, making these ratios potential chemical clocks for stars  \citep[e.g.][]{Spina2016,2018MNRAS.474.2580S,2016A&A...593A..65N,2020A&A...633L...9J,Viscasillas2022, Owusu2024,Molero2025}. Many studies have been dedicated to investigations of [Y/Mg] \citep[e.g.][]{2015A&A...579A..52N,2016A&A...593A..65N,2016A&A...590A..32T,2017A&A...608A.112N,2020A&A...639A.127C,Tautvaisiene2021}. 

The second mechanism relies on stellar evolution and internal mixing. Specifically, the photospheric ratio of carbon to nitrogen ([C/N]) in evolved stars changes as the material processed by the CNO cycle is brought to the surface during the first dredge-up (1DUP; \citep{Iben65, Charbonnel94, Boothroyd99}). Since the depth of mixing and the main-sequence lifetime are strictly mass dependent, the [C/N] ratio in giants serves as a potent proxy for stellar mass and consequently age. The use of [C/N] as an age indicator was theoretically posited by \citet{salaris2015} for RGB stars and by \citet{2017A&A...601A..27L} for red clump (RC) stars. Although observational studies have confirmed the utility of [C/N] as an age indicator 
\citep[e.g.][]{Martig16,hasselquist19,2019A&A...629A..62C,2022AJ....163..229S,Spoo2025,Roberts25,2025A&A...703A...4T,Lu2026}, calibrations remain uncertain, particularly at the young and old age extremes, where standard models struggle to capture the full extent of mixing efficiencies. In these regimes, non-standard processes such as rotation- and thermohaline-induced mixing become highly influential and are currently less constrained \citep[e.g.][]{gratton20, eggleton2008, cantiello2010, denissenkov2010, Charbonnel10, Lagarde12, Lagarde2019, Aguilera23, Lagarde2024}. For instance, rapid rotation in young massive main-sequence stars can significantly alter the internal chemical profiles and cause additional abundance alterations during the 1DUP, leading to [C/N] ratios that deviate from standard predictions. On the other hand, in older lower-mass giants, thermohaline mixing occurring after the RGB luminosity bump can induce further surface abundance changes, complicating the interpretation of [C/N] as a pure age proxy. To disentangle these mass- and age-dependent mixing efficiencies, a large sample of stars with independently known high-precision ages, such as members of open clusters, is essential.   

The Stellar Population Astrophysics (SPA) programme is a large observing programme conducted at the Telescopio Nazionale Galileo (TNG) designed to investigate the age-resolved chemical properties of the Milky Way disc \citep[e.g.][]{zhang21, zhang2022, seshashayana2024}. The programme exploits the exceptionally high-resolution spectroscopic capabilities of the HARPS-N and GIANO-B instruments, which together provide continuous coverage from the optical to the near-infrared, extending to the $K$ band \citep{claudi17}. 

Within the SPA framework, a large sample of 33 open clusters was recently observed and non-local thermodynamic (NLTE) atmospheric parameters as well as abundances of ten chemical elements were determined from the HARPS-N spectra of giant stars in them \citep{2025A&A...701A.289D}. In this study, we expand previous analysis by determining the carbon, nitrogen, and oxygen abundances and by investigating the relation of [C/N] with age. Using the precise isochrone ages of these clusters, we aim to refine the [C/N]--age calibration and investigate the dependence of the C/N\footnote{C/N is defined as $N_C$/$N_N$, where $N_C$ and $N_N$ are the number densities of carbon and nitrogen nuclei. The [C/N], on the other hand, is the solar-normalised logarithmic ratio: [C/N] = log($N_C$/$N_N$)$_\mathrm{star}$ -- log($N_C$/$N_N$)$_\mathrm{Sun}$. } ratio on turn-off mass across different evolutionary stages. This approach allowed us to empirically constrain the mixing processes and the validity of the [C/N] clock across a wide age range. The paper is organised as follows: In Section~\ref{sample and method} we detail the stellar sample and the methodology used to determine atmospheric parameters and chemical abundances. Section~\ref{results} presents our results and is followed by a discussion, and the conclusions are presented in Section~\ref{sec:summary_conclusions}.

\section{Stellar sample and method of analysis} 
\label{sample and method}

\subsection{Cluster member sample selection}

Our analysis focuses on evolved stars that reside on the RGB and in the RC. Our sample was drawn from likely members of open clusters as identified by \citet{2018A&A...615A..49C, 2020A&A...640A...1C} using the $Gaia$  Data Release~2 (DR2) astrometry and photometry \citep{2018A&A...616A...1G}. 
This sample was further reviewed and refined in the recent SPA study by \cite{2025A&A...701A.289D} to ensure the reliability of membership.

\begin{figure*}
    \centering
    \includegraphics[width=\textwidth]{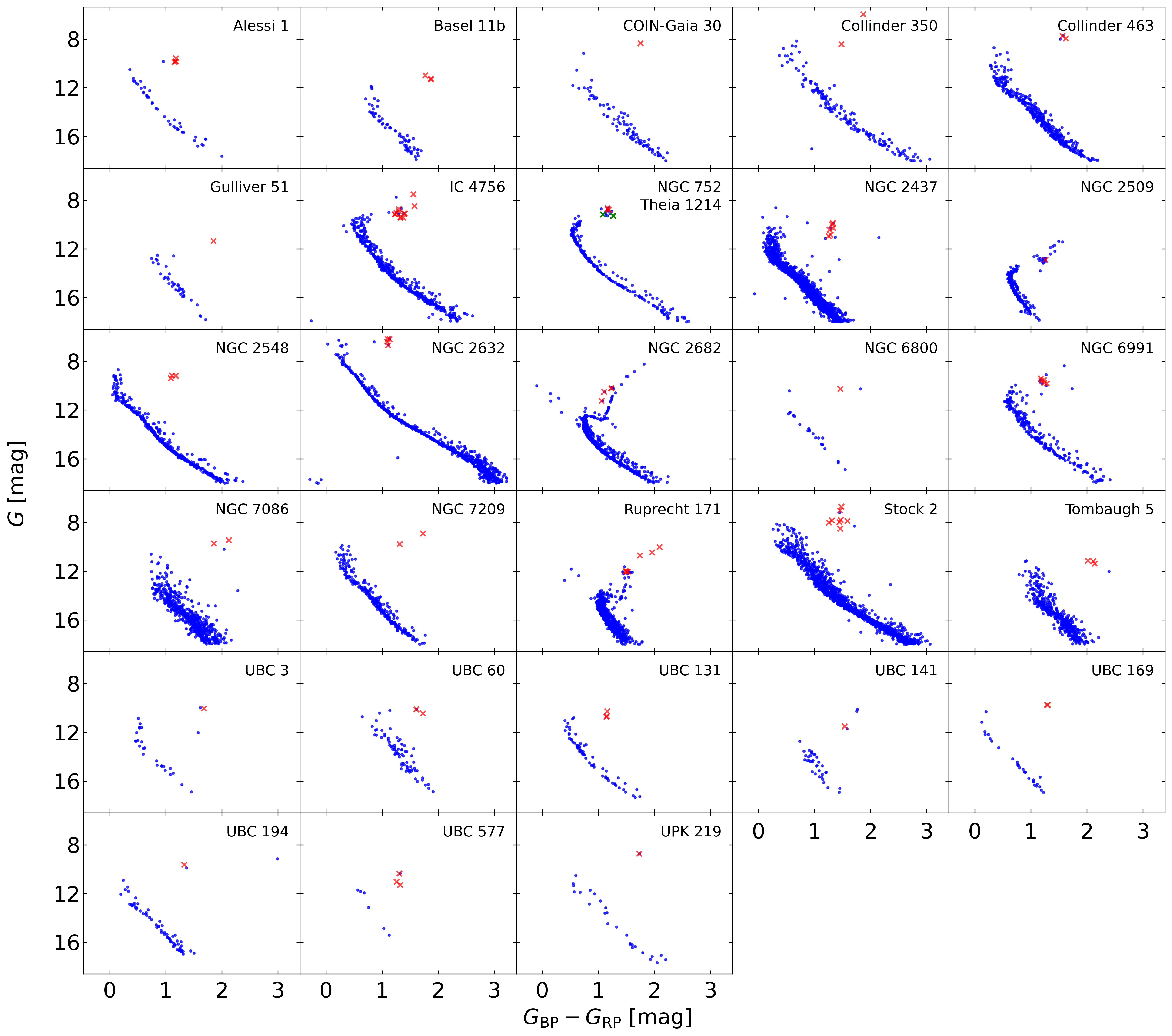}
    \caption{Colour–magnitude diagrams of open cluster stars using Gaia DR3 photometry. The axes represent the $Gaia$ $(G_{BP} - G_{RP})$ colour and absolute $G$ magnitude. The blue dots represent high-probability members from \citep{2020A&A...640A...1C}, while the red crosses indicate the stars analysed in this work. The Theia\,1214 stars are plotted alongside NGC\,752 (indicated by the green crosses) due to their suggested association with NGC\,752 \citep{2025A&A...701A.289D}.  }
    \label{fig:CMDs}
\end{figure*}

We analysed high-resolution optical spectra previously investigated by \cite{2025A&A...701A.289D} and added new observations of six stars (NGC\,2437\_9, NGC\,2632\_HD\,73794 and HD\,73710, NGC\,2682\_1, NGC\,2682\_2, UBC\,60\_4). Spectra were acquired with the HARPS-N spectrograph ($R \approx 115\,000$) with a wavelength range from 3800 to 6900~{\AA}. 
The exposure times were adjusted to account for differences in the target brightness and observing conditions. Following the observations, the spectra were processed with the automated HARPS-N reduction pipeline. As part of this procedure, the barycentric velocity correction was applied, the instrumental response was calibrated, wavelength solutions were derived, and one-dimensional spectra were extracted. For stars observed multiple times, the individual exposures were combined into a single spectrum prior to abundance analysis. 

The colour-magnitude diagrams of our sample are provided in Fig.~\ref{fig:CMDs}, for which we utilised photometry from \textit{Gaia} DR3 and adopted memberships based on \citet{2020A&A...640A...1C}. The fundamental parameters of open clusters studied in this work are listed in  Table~\ref{tab:clusters}. We provide the number of stars investigated in each cluster, the mean metallicity of the cluster stars investigated and the scatter, the age from \cite{2020A&A...640A...1C}, the mean Galactocentric radius ($R_{\rm mean}$), and the maximum height above the Galactic plane ($z_{\rm max}$), computed in this work (Subsect.~\ref{Locations}). 

\begin{table*}
\centering
\caption{Basic parameters of the investigated objects. }
\label{tab:clusters}
\begin{tabular}{lcccccccc}
\hline\hline
Cluster name & Alternative names & N & [Fe/H]\tablefootmark{a} & log(Age[yr])\tablefootmark{b} & Turn-off mass & $R_{\rm mean}$ & $|z_{\rm max}|$ \\
             &             &             &        &  & $M_\odot$ & kpc & kpc \\
\hline
Alessi\,1        & Casado-Alessi\,1      & 4 & $-0.01$ $\pm$ 0.01 & 9.16  & 1.62 & 8.13 & 0.24 \\
Basel\,11b  & FSR\,887           & 3 & 0.04 $\pm$ 0.03 & 8.36  & 3.20 & 9.02 & 0.07  \\
COIN-Gaia\,30    & –          & 1 & $-0.04$  & 8.41  & 3.08 & 8.84 & 0.18  \\
Collinder\,350    & –             & 2 & $-0.04 \pm$ 0.12 & 8.77  & 2.23 & 7.71 & 0.19  \\
Collinder\,463    & –             & 2 & $-0.06 \pm$ 0.02 & 8.06  & 4.26 & 9.00 & 0.18  \\
Gulliver\,51    & –             & 1 & $-0.17 $ & 8.56  & 2.64 & 9.25 & 0.14  \\
IC\,4756    & Collinder\,386, Melotte\,210 & 12 & 0.00 $\pm$ 0.04 & 9.11  & 1.70 & 7.25 & 0.08  \\
NGC\,2437    & M\,46, Melotte\,75 & 6 & $-0.02 \pm$ 0.08 & 8.48  & 2.89 & 8.69 & 0.30  \\
NGC\,2509    & Melotte\,81, Collinder\,171 & 3 & 0.21 $\pm$ 0.01 & 9.18  & 1.60 & 8.68 & 0.30  \\
NGC\,2548    & M 48, Melotte 85   & 3 & 0.05 $\pm$ 0.06 & 8.59  & 2.64 & 9.04 & 0.28  \\
NGC\,2632    & M\,44, Praesepe, Melotte\,88  & 4 & 0.14 $\pm$ 0.04 & 8.83  & 2.18 & 7.90 & 0.13  \\
NGC\,2682    & M\,67             & 4 & 0.04 $\pm$ 0.03 & 9.63  & 1.17 & 8.39 & 0.60  \\
NGC\,6800    & –             & 1 & 0.11 & 8.61  & 2.64 & 7.43 & 0.10  \\
NGC\,6991    & –             & 5 & 0.04 $\pm$ 0.06 & 9.19  & 1.60 & 8.05 & 0.15  \\
NGC\,7086    & Collinder\,437  & 2 & $-0.08 \pm$ 0.03 & 8.29  & 3.44 & 8.09 & 0.11   \\
NGC\,7209    & Melotte\,238, Collinder 444 & 2 & $-0.03 \pm$ 0.07 & 8.63  & 2.53 & 7.90 & 0.15  \\
NGC\,752   & Melotte\,12  & 3 & 0.08 $\pm$ 0.02 & 9.07  & 1.76 & 8.01 & 0.23  \\
Ruprecht\,171    & –             & 7 & $-0.07 \pm$ 0.06 & 9.44  & 1.29 & 8.24 & 0.65  \\
Stock\,2    & –             & 8 & $-0.02 \pm$ 0.06 & 8.60  & 2.60 & 8.09 & 0.10  \\
Tombaugh\,5    & –             & 3 & 0.02 $\pm$ 0.07 & 8.27  & 3.55 & 9.04 & 0.18  \\
UBC\,3    & Alessi\,161   & 1 & 0.00  & 8.10  & 4.15 & 6.94 & 0.19  \\
UBC\,60   & COIN-Gaia\,11     & 2 & 0.18 $\pm$ 0.05 & 8.90  & 2.04 & 8.27 & 0.08  \\
UBC\,131    & Alessi\,116, UPK 84  & 3 & 0.05 $\pm$ 0.03 & 9.00  & 1.85 & 8.48 & 0.22  \\
UBC\,141    & –             & 1 & $-0.02 $ & 9.32  & 1.45 & 7.62 & 0.27  \\
UBC\,169    & –             & 2 & 0.06 $\pm$ 0.03 & 8.47  & 2.96 & 8.37 & 0.21  \\
UBC\,194    & –             & 1 & 0.06 & 8.36  & 3.26 & 8.57 & 0.24  \\
UBC\,577    & Alessi\,191      & 3 & $-0.05 \pm$ 0.02 & 9.44  & 1.30 & 7.79 & 0.32  \\
UPK\,219    & –             & 1 & 0.07 & 8.17  & 3.88 & 9.15 & 0.08  \\
\\
Theia\,1214\_1\tablefootmark{c}    &   &  1 & $0.14$ & 8.69\tablefootmark{d}  & 2.78\tablefootmark{d} &  7.95  & 0.23 \\
Theia\,1214\_3\tablefootmark{c}    &   &  1 & $-0.25 $  &  9.83\tablefootmark{d}  & 1.08\tablefootmark{d} & 9.31  & 0.62\\
\hline
\end{tabular}
\tablefoot{\\
\tablefoottext{a}{The mean [Fe/H] of individual measured values for stars of each cluster in \cite{2025A&A...701A.289D}.} \\
\tablefoottext{b}{Ages are taken from \citet{2020A&A...640A...1C}.}\\
\tablefoottext{c}{This association is inferred to be a part of NGC\,752, as a tidal tail \citep{2025A&A...701A.289D}, but we treat it as a separate entity in this study.} \\
\tablefoottext{d}{For the Theia\,1214 stars, the age and mass were derived in this study.  }
}
\end{table*}

\subsection{Atmospheric parameters}

The main atmospheric parameters -- effective temperature ($T_{\rm eff}$), surface gravity ($\log g$), metallicity ([Fe/H]), and microturbulent velocity ($\xi$) -- were adopted from \citet{2025A&A...701A.289D} and determined using the same method for the newly observed stars (Table~\ref{tab:stellar_params}). We summarise the methodology here. 

The parameters were determined using an equivalent width method. Equivalent widths were measured for 165 neutral and 20 ionised iron lines using the \texttt{smhr} tool \cite{Casey2014} and the \texttt{stellardiff} tool.\footnote{https://github.com/andycasey/stellardiff}  Lines stronger than $200$ m\AA\  and with an error greater than 10~m\AA\ were excluded. 

To determine the main atmospheric parameters, the \texttt{LOTUS} pipeline (non-LTE optimization tool; \citealt{Li2023}) was used. This code interpolates a generalised curve of growth in a grid of theoretical equivalent widths calculated under NLTE conditions. The final parameters were obtained through a global minimisation procedure that enforces both excitation and ionisation equilibrium conditions. Parameter uncertainties were robustly estimated using a Markov chain Monte Carlo algorithm. This NLTE framework was adopted to account for departures from local thermodynamic equilibrium in the outer atmospheres of cool giant stars.

\begin{table*}[]
\centering
\caption{Atmospheric parameters determined in this work.}
\label{tab:stellar_params}
\begin{tabular}{lccccc}
\hline\hline
Star & Gaia DR3 ID & $T_{\rm eff}$ (K) & $\log g$  & [Fe/H]  & $\xi$ (km\,s$^{-1}$) \\
\hline
NGC\,2437\_9 & 3029207006148017664  & 4821 $\pm$ 38 & 2.47 $\pm$ 0.10 & 0.01 $\pm$ 0.01 & 1.59 $\pm$ 0.02 \\
NGC\,2632\_HD73794 & 661396754238802816  & 4898 $\pm$ 34 & 2.77 $\pm$ 0.10 & 0.08 $\pm$ 0.01 & 1.39 $\pm$ 0.02 \\
NGC\,2632\_HD73710 & 661297080936069632  & 4827 $\pm$ 35 & 2.66 $\pm$ 0.09 & 0.15 $\pm$ 0.01 & 1.43 $\pm$ 0.03 \\
NGC\,2682\_1 & 604917728138508160  & 5089 $\pm$ 34 & 3.16 $\pm$ 0.08 & $-$0.01 $\pm$ 0.01 & 1.12 $\pm$ 0.02 \\
NGC\,2682\_2 & 604904950611554432 & 5087 $\pm$ 31 & 3.32 $\pm$ 0.09 & 0.07 $\pm$ 0.01 & 1.98 $\pm$ 0.02 \\
UBC\,60\_4 & 179632166729908992 & 5047 $\pm$ 36 & 3.09 $\pm$ 0.09 & 0.13 $\pm$ 0.01 & 1.37 $\pm$ 0.02 \\
\hline
\end{tabular}
\end{table*}

\subsection{Abundances of carbon, nitrogen, and oxygen}

We used the Turbospectrum (v. 2019) code \citep{alvarez98, Plez12} with MARCS atmosphere models \citep{Gustafsson08} to generate synthetic spectra. We adopted the solar abundances by \cite{Grevesse2007}, and the atomic and molecular line list previously used in the $Gaia$-ESO survey (see \citealt{Heiter21} and references therein). 

In cool stellar atmospheres ($T_{\rm eff} \lesssim 5000$ K), the determination of carbon, nitrogen, and oxygen abundances is strongly coupled due to the formation of molecular equilibrium species, most notably CO and CN \citep{Lambert84, Ryabchikova22}. Even at solar temperatures, \cite{Amarsi21} demonstrates that accurate C, N, and O abundances require a simultaneous solution of the molecular equilibrium. In the cooler regime studied here ($T_{\rm eff} \lesssim 5000$ K), this coupling intensifies as the high dissociation energy of CO and CN effectively locks the minority element (carbon or oxygen) into the molecular phase, leaving only the excess abundance available for other chemical associations. Consequently, a variation in the adopted oxygen abundance directly modulates the determined abundance of free carbon, which in turn governs the formation rates of other diagnostics such as the CN and C$_2$ Swan bands \citep{Ryde09}. Therefore, we adopted an iterative process to derive C, N, and O abundances that satisfy the molecular equilibrium equations consistent with the observed spectral features.

First, the abundance of oxygen was derived from the forbidden [\ion{O}{I}] line at 6300.3~{\AA}. This line is largely insensitive to NLTE effects and has been shown to yield consistent abundances in both 1D and 3D model atmospheres \citep{Amarsi16}. For Basel\,11b\_3, the oxygen line was heavily blended by the telluric line. For this star, we used the averaged oxygen abundance from the other two stars, which have very similar oxygen abundances (0.17 and 0.18~dex). With the oxygen abundance constrained, carbon was derived from the C$_2$ Swan band heads at 5135 and 5635~\AA. These molecular features are preferred to atomic lines for cool stars, as they provide results consistent with [\ion{C}{I}] diagnostics while remaining robust against NLTE deviations \citep{Ryabchikova22}. Finally, nitrogen abundances were inferred from the features of $^{12}$C$^{14}$N in the 6470--6490~{\AA} region.

Given the interdependence of the molecular features described above, errors in one element inevitably propagate to the others. To quantify these systematic uncertainties, we performed a sensitivity analysis on a representative star, selected for having atmospheric parameters closest to the median of the full sample. We prioritised the median over the mean to ensure error estimates that are more representative of the typical star in our dataset were provided. We perturbed the input abundances of C, N, and O individually by $\pm 0.10$ dex and re-derived the equilibrium abundances of the remaining species. The resulting deviations are summarised in Table \ref{tab:CNO_err}. We observed a significant anti-correlation between carbon and nitrogen ($\mp 0.11$ dex). This behaviour is expected, as an overestimation of the carbon abundance enhances the formation efficiency of CN, which requires a compensatory decrease in nitrogen to reproduce the observed CN line depths. Conversely, carbon correlates positively with oxygen ($\pm 0.03$ dex) due to the association of CO, which dominates the carbon budget in oxygen-rich atmospheres. 

We also evaluated how the median uncertainties of the atmospheric parameters ($\pm 35$~K in $T_{\rm eff}$, $\pm 0.09$~dex in log\,$g$, 0.01~dex in [Fe/H], and 0.02~km\,s$^{-1}$ in $v_{\rm t}$) could change the CNO abundances. The influence does not exceed $\pm 0.01$ -- only the uncertainty of log\,$g$ can cause alterations up to  $\pm 0.04$~dex in the abundance of oxygen.

\begin{table}
\centering
\caption{Sensitivity of derived CNO abundances to perturbations in assumed input abundances for the star NGC\,2437\_5.}
\label{tab:CNO_err}
\begin{tabular}{lccc}
\hline\hline
			& $\Delta$[C/H] & $\Delta$[N/H] & $\Delta$[O/H] \\
            &     $\pm 0.10$ dex        &    $\pm 0.10$ dex         & $\pm 0.10$ dex \\
\hline
$\Delta$[C/H]  & ---  &  $\pm 0.01$   & $\pm 0.03$ \\
$\Delta$[N/H]  & $\mp0.11$  &  ---   & $\pm 0.06$  \\
$\Delta$[O/H]  &  $\pm 0.01$ &  $\pm 0.00$   & ---  \\
$\Delta$C/N & $\pm 0.23$   & $\mp 0.23$   & ---  \\
\hline
\end{tabular}
\end{table}

\subsection{Determining the evolutionary stage of stars}

To investigate potential age-dependent trends in chemical abundances, we determined the evolutionary stage of each target by analysing its position in the Kiel diagram ($\log g$ versus $T_{\rm eff}$). We individually compared our derived atmospheric parameters with PARSEC isochrones  \citep{2012MNRAS.427..127B} for every cluster. For this comparison, we adopted the cluster ages from \citet{2020A&A...640A...1C}, which were derived using artificial neural networks applied to \textit{Gaia} DR2 photometry \citep{2018A&A...616A...1G}. The metallicity for each isochrone was fixed to the mean cluster [Fe/H] determined in this work.
 
We classified the stars into three evolutionary groups: first-ascent red giants, situated below the luminosity bump, burning hydrogen in a shell (abbreviated RGB); stars above the RGB luminosity bump (abbreviated RGB\,AB); and core helium burning RC stars. The distinction between pre- and post-bump stars is physically significant. The RGB luminosity bump corresponds to a temporary hydrostatic equilibrium established when the H-burning shell encounters the chemical discontinuity left by the maximum penetration of the convective envelope during the 1DUP. This interaction causes a temporary decrease and a subsequent increase in luminosity, creating an overdensity in the HR diagram.  

Although a majority of the stars were classified purely by their location relative to the isochrones, there is a known degeneracy between the RC and the RGB in the $\log g$--$T_{\rm eff}$ plane. In ambiguous cases, we utilised carbon-to-nitrogen ratios (C/N) as a secondary discriminant. As a result of deep mixing events, RC stars typically exhibit lower [C/N] and C/N ratios compared to lower RGB stars of similar metallicity \citep[c.f.][and references therein]{gratton20, Lagarde12, 2025A&A...703A...4T}. We effectively identified RGB\,AB stars (post-bump giants) in only three of the open clusters in our sample.

\subsection{Locations of the open clusters in the Galaxy}
\label{Locations}

To investigate abundance gradients across the Galactic disc, we computed the orbital parameters for our sample of open clusters. We utilised the Python package \texttt{galpy}\footnote{\url{http://github.com/jobovy/galpy}} \citep{Bovy15} to integrate the orbits. The input data were constructed using the celestial coordinates ($\alpha, \delta$), proper motions ($\mu_{\alpha*}, \mu_{\delta}$), and parallaxes ($\varpi$) from \textit{Gaia}\,DR3 using \texttt{Astropy} \citep{Astropy22} and the \texttt{pyia} package \citep{pyia:v1_3}.

The orbits were integrated within the default static axis symmetric potential \texttt{MWPotential2014}, which models the Milky Way's bulge, disc, and dark matter halo. We adopted a Galactocentric distance to the Sun of $R_\odot = 8$\,kpc and a vertical height above the plane of $z_\odot = 20.8$\,pc \citep{Bennett18}. The circular velocity at the solar radius was set to $V_c(R_\odot) = 220$\,km\,s$^{-1}$ \citep{Kerr86}, and the solar motion values are (U, V, W) = (11.1, 12.24, 7.25)~km\,s$^{-1}$ \citep{Schoenrich10}.

\begin{figure}
    \centering
    \includegraphics[width=1.0\columnwidth]{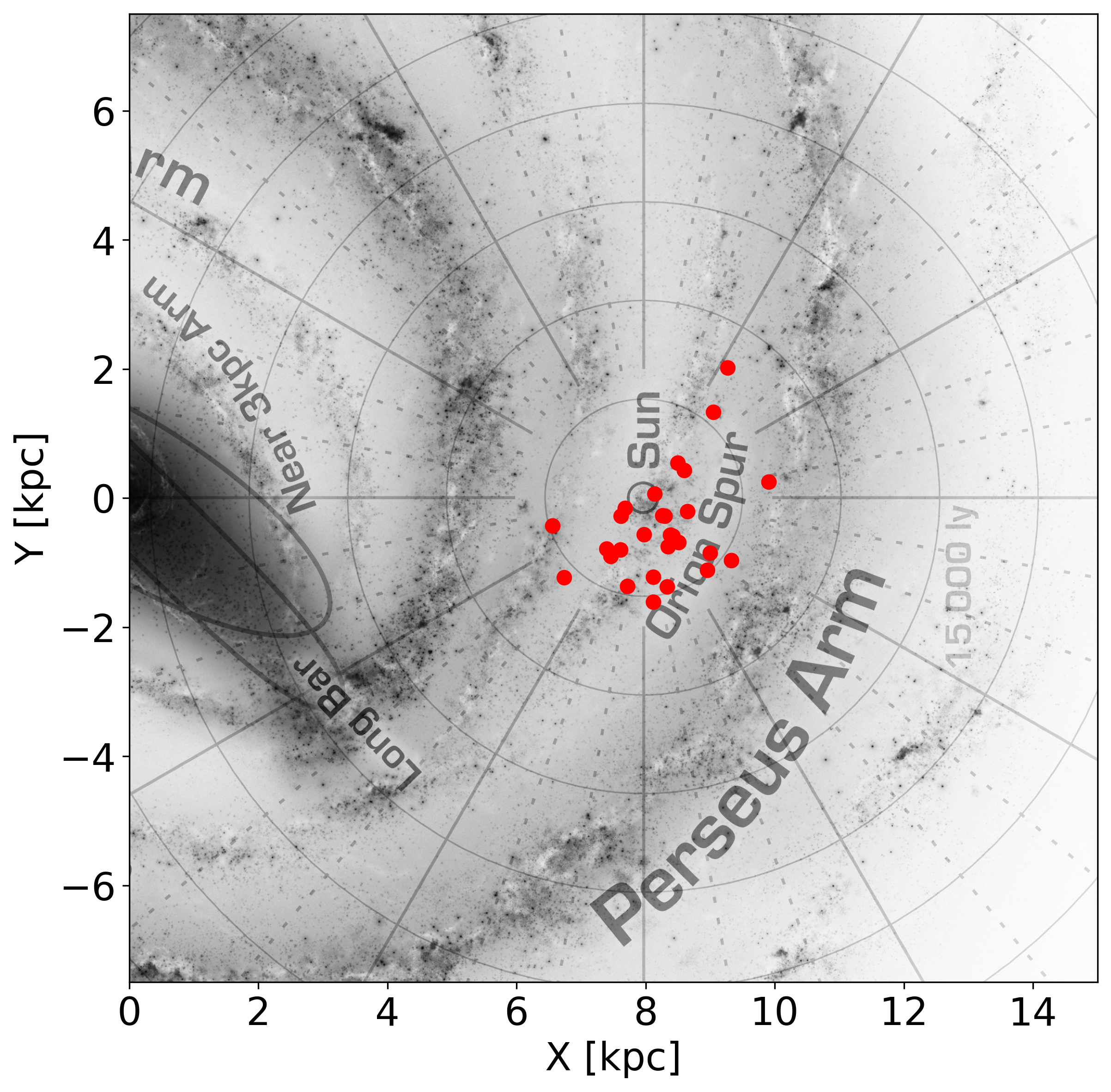}
    \caption{Locations of the investigated open clusters in the Galaxy.}
    \label{fig:MW_clusters}
\end{figure}

For each cluster, the mean Galactocentric distance ($R_{\rm mean}$) was calculated as the arithmetic mean of the apogalactic and perigalactic distances derived from the orbital integration. The locations of the investigated open clusters in the Galaxy are displayed in Fig.~\ref{fig:MW_clusters}.

\subsection{Overview of the sample}
\begin{figure}
    \centering
    \includegraphics[width=0.49\columnwidth]{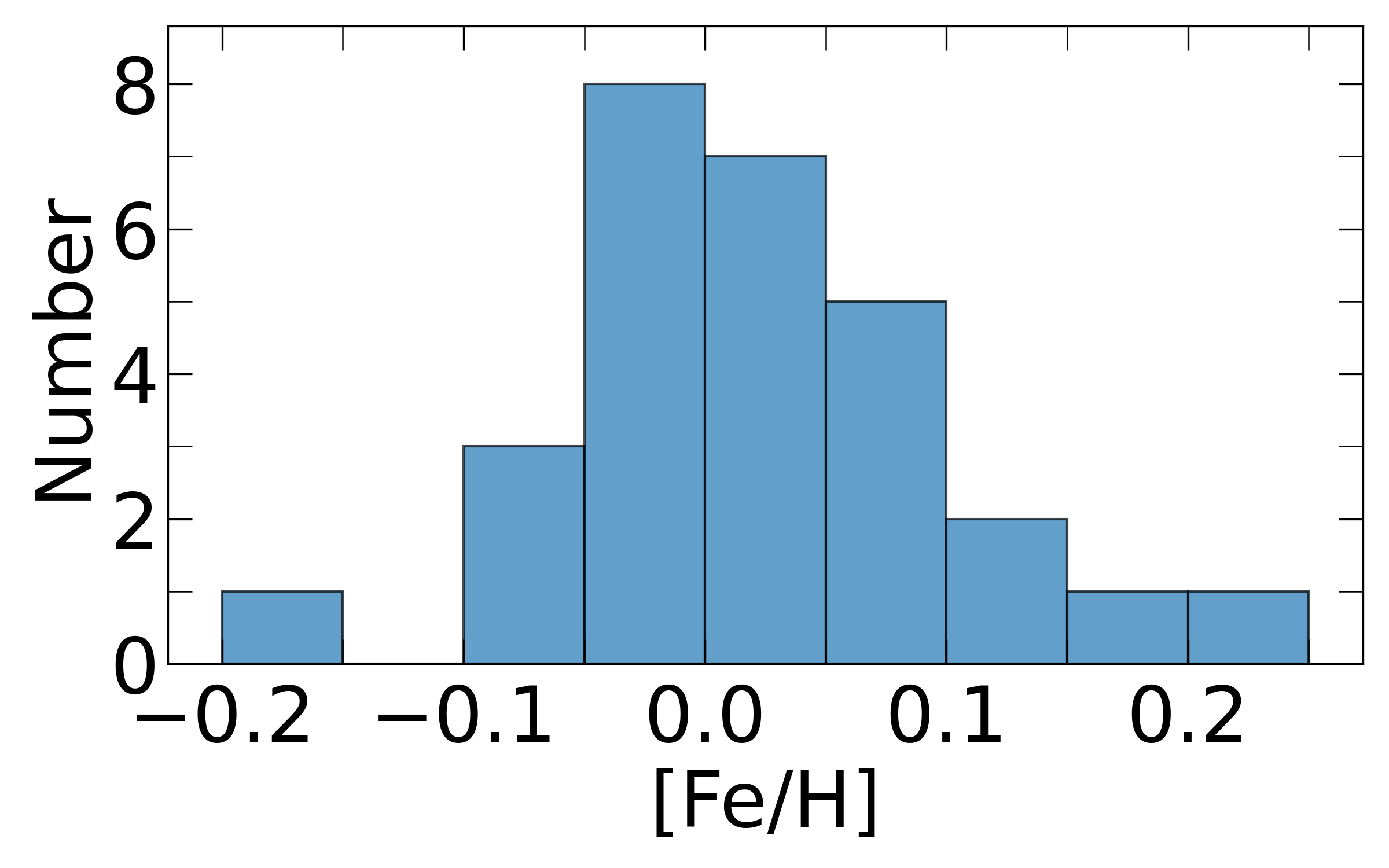}
    \includegraphics[width=0.49\columnwidth]{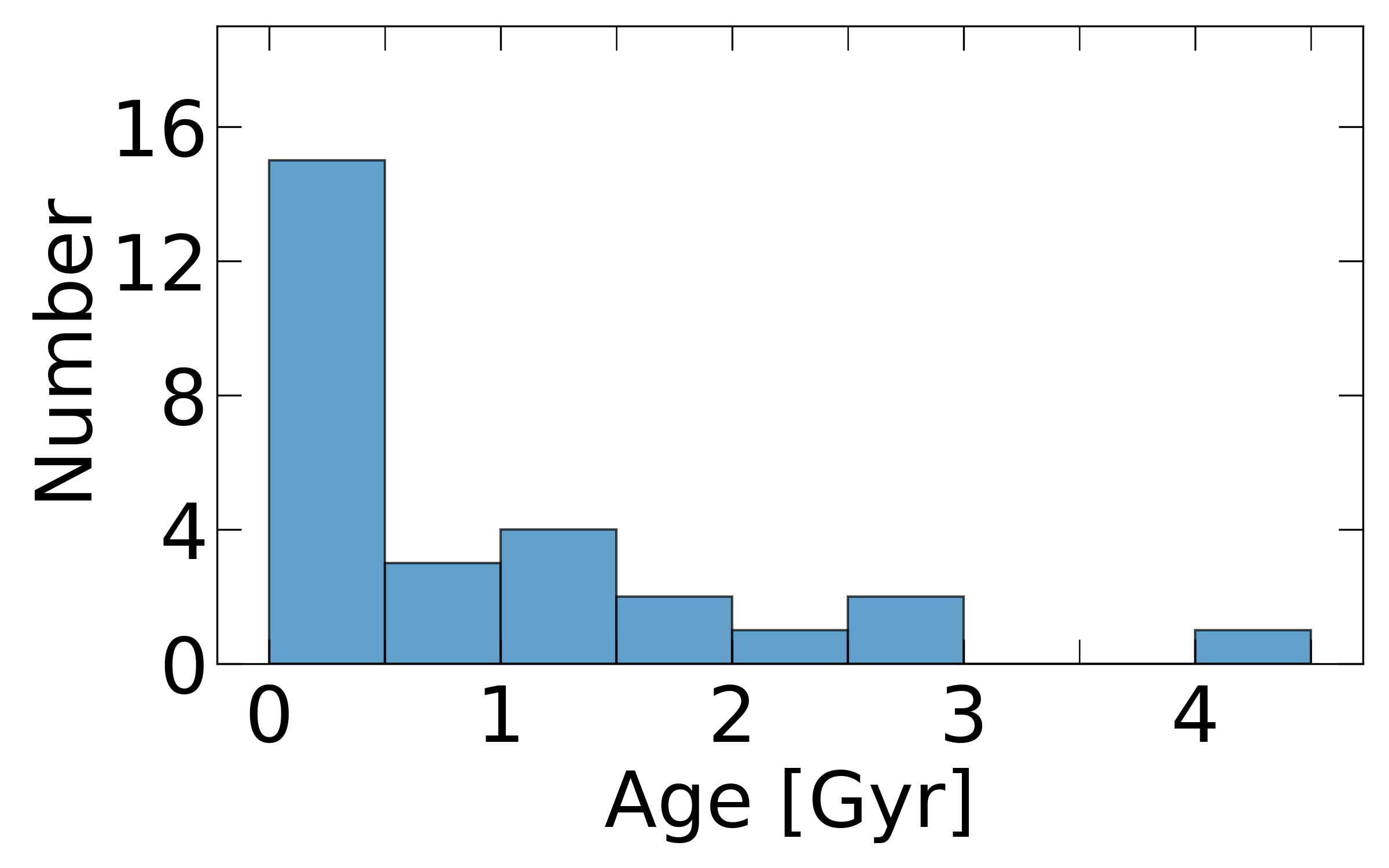}
    \includegraphics[width=0.49\columnwidth]{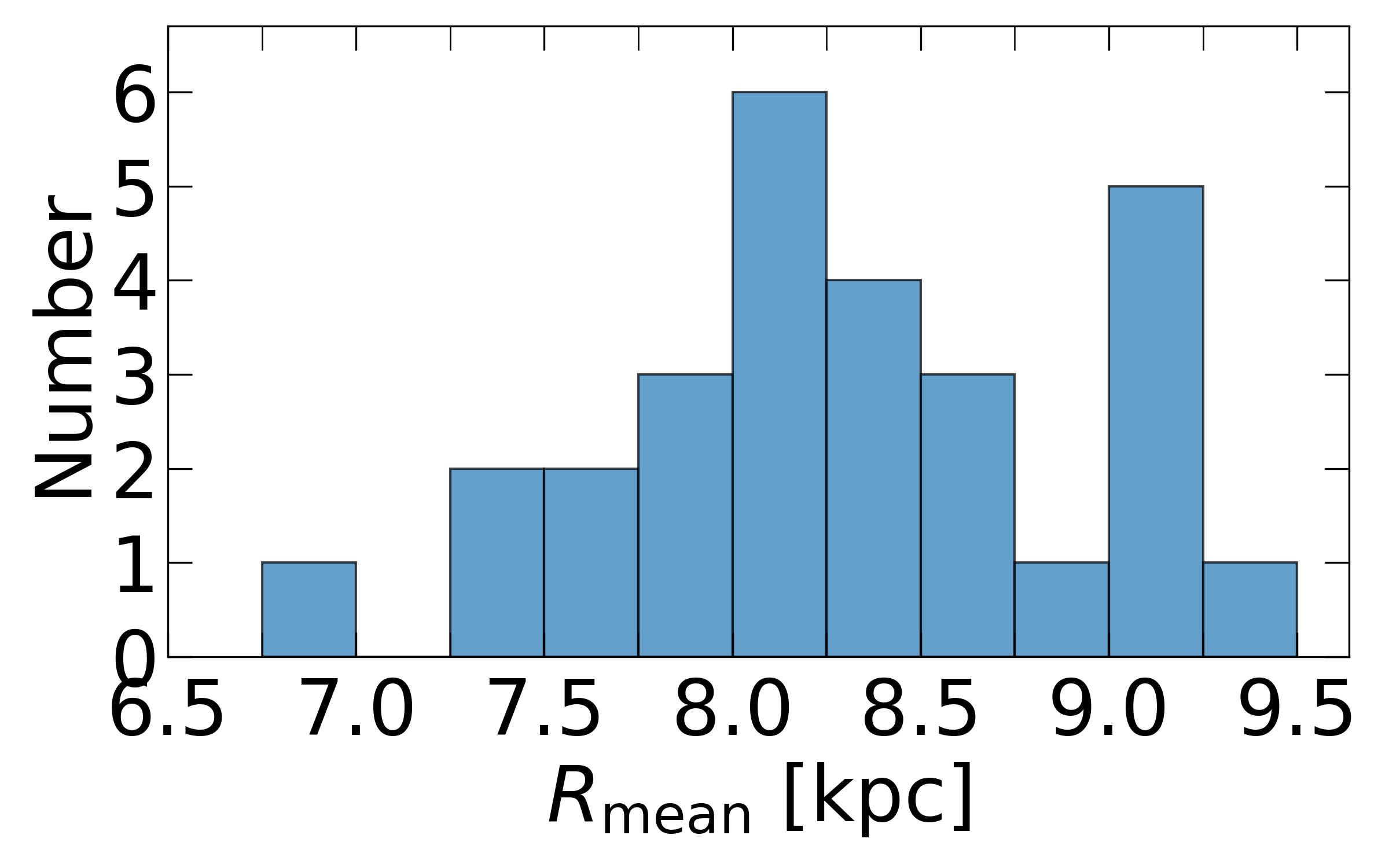}
        \includegraphics[width=0.49\columnwidth]{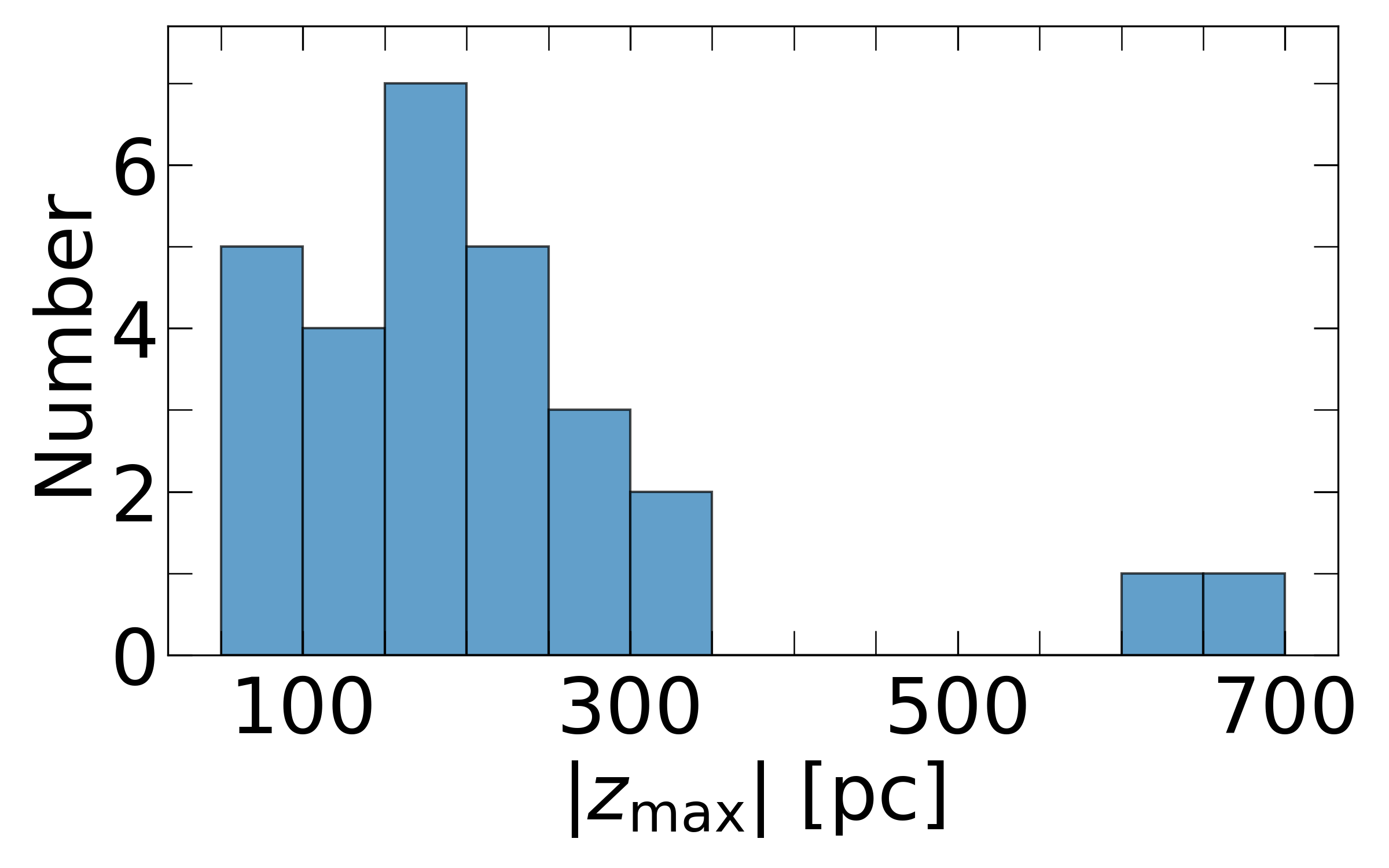}
    \caption{Histograms of metallicities, ages, mean galactocentric distances, and maximal distances from the Galactic plane for the investigated open clusters with measured CNO abundances in this work. }
    \label{fig:histograms}
\end{figure}

The final sample analysed in this work comprises 88 evolved stars that belong to 28 open clusters. The distribution of the sample in terms of cluster age, metallicity ([Fe/H]), mean Galactocentric radius ($R_{\rm mean}$), and maximum vertical excursion from the plane ($Z_{\rm max}$) is presented in Fig.~\ref{fig:histograms}. 

As shown in the histograms, the sample is characteristic of the Galactic thin disc population. It is predominantly composed of young to intermediate-age clusters clustering tightly around solar metallicity. Spatially, the sample covers a range of Galactocentric radii from approximately $R_{\rm mean} \approx 7$ to 9.5\,kpc. However, the distribution exhibits a notable asymmetry, with a higher density of targets located in the outer disc ($R_{\rm mean} > R_\odot$) compared to the inner disc.

\begin{table*}
\caption{Carbon, nitrogen, and oxygen abundances and evolutionary phases for the analysed stars. }
\label{tab:cno_abundances}
\centering
\begin{tabular}{lcccccccccc}
\hline\hline
Star & $Gaia$ DR\,3 ID &
[C/H] & $\sigma_{\mathrm{C}}$ & n &
[N/H] & $\sigma_{\mathrm{N}}$ & n &
[O/H] & $\sigma_{\mathrm{O}}$ &
Evol. \\

\hline
Alessi\,1\_2 & 402506369136008832 &$-0.24$ & 0.01 & 2 & 0.24 & 0.04 & 2 & 0.00 & 0.04 & RC\\
Alessi\,1\_3 & 402505991180022528 &$-0.26$ & 0.02 & 2 & 0.26 & 0.04 & 2 & 0.00 & 0.04 & RC\\
Alessi\,1\_5 & 402867593065772288 &$-0.24$ & 0.02 & 2 & 0.24 & 0.04 & 2 & 0.03  & 0.03 &  RC\\
Alessi\,1\_6 & 402880684126058880 &$-0.24$ & 0.02 & 2 & 0.22 & 0.06 & 3 & 0.04  & 0.03 &  RC\\
Basel\,11b\_1 & 3424056131485038592 &$-0.21$ & 0.02 & 2 & 0.44 & 0.05 & 3 & 0.17 & 0.03 & RGB \\
Basel\,11b\_2 & 3424055921028900736 &$-0.17$ & 0.02 & 1 & 0.45 & 0.07 & 3 & 0.18  & 0.04 & RGB \\
Basel\,11b\_3 & 3424057540234289408 &$-0.06$ & 0.02 & 2 & 0.45 & 0.06 & 3 & --  & -- & RGB \\
COIN-Gaia\,30\_1 & 532533682228608384 &$-0.22$ & 0.02 & 2 & 0.35 & 0.06 & 3 & 0.10 & 0.04 & RGB \\
Collinder\,350\_1 & 4372743213795720704 &$-0.30$ & 0.01 & 2 & 0.42 & 0.02 & 3 & 0.05 & 0.00 & RGB AB\\
... & ... & ... & ...& ... & ... & ... & ... & ... & ... & ... \\
\hline
\end{tabular}
\tablefoot{Only an excerpt of the table is displayed here. The full table is accessible through the CDS. The evolutionary stages of stars: RGB – red giant branch below the luminosity bump, RC – red clump, RGB\,AB – red giant above the RGB luminosity bump. n -- number of lines  investigated. }
\end{table*}

\section{Results and discussion}
\label{results}

The derived C, N, and O abundances along with the evolutionary stages determined for each target are compiled in Table~\ref{tab:cno_abundances}. In this section, we first interpret these abundance patterns in the context of theoretical stellar evolution models, focusing on the chemical signatures of the 1DUP and subsequent mixing events. We then evaluate the efficacy of the [C/N] ratio as an empirical age diagnostic (chemical clock), specifically analysing how the abundance-age correlation varies across the distinct evolutionary phases (RGB versus RC) identified in our sample.

\subsection{Comparison C/N ratios with theoretical evolutionary models}\label{model_comparison}

 Standard stellar evolution theory attributes the primary alteration of surface carbon and nitrogen abundances in low- and intermediate-mass stars to the 1DUP event. As the star ascends the RGB, the convective envelope deepens into regions previously processed by the CNO cycle, bringing nuclear-processed material to the surface. This process is expected to decrease the surface abundance of $^{12}$C, increase $^{14}$N, and lower the $^{12}$C/$^{13}$C isotopic ratio to values of approximately 20--30, depending on stellar mass and metallicity (e.g. \citealt{Iben65}).

However, observational evidence consistently demonstrates that stars brighter than the RGB `luminosity bump' (i.e. the RGB\,AB and RC phases) exhibit chemical signatures that deviate from standard 1DUP predictions. These stars frequently display $^{12}$C/$^{13}$C ratios significantly lower than standard 1DUP predictions (often reaching values below ten), along with  varying degrees of Li depletion and reduction of C/N ratio \citep[e.g.][and references therein]{Gilroy89, Smiljanic09,motta17,Lagarde2024}. These discrepancies indicate that non-canonical transport mechanisms -- `extra mixing' -- must be active during the post-1DUP phase.

The primary physical mechanisms proposed to explain this extra mixing are thermohaline instability and rotation-induced mixing. Thermohaline mixing is a double-diffusive instability triggered by a mean molecular weight inversion created by the $^3\text{He}(^3\text{He}, 2p)^4\text{He}$ reaction in the hydrogen-burning shell \citep{Charbonnel07}. This instability efficiently transports material between the burning shell and the convective envelope, further processing carbon into nitrogen and significantly reducing the $^{12}$C/$^{13}$C ratio. Additionally, rotation-induced mixing, driven by meridional circulation and shear turbulence, smooths abundance profiles during the main-sequence phase and causes modifications of surface compositions during the RGB ascent \citep{Charbonnel10}. 

\begin{figure}
    \centering
    \includegraphics[width=1.1\columnwidth]{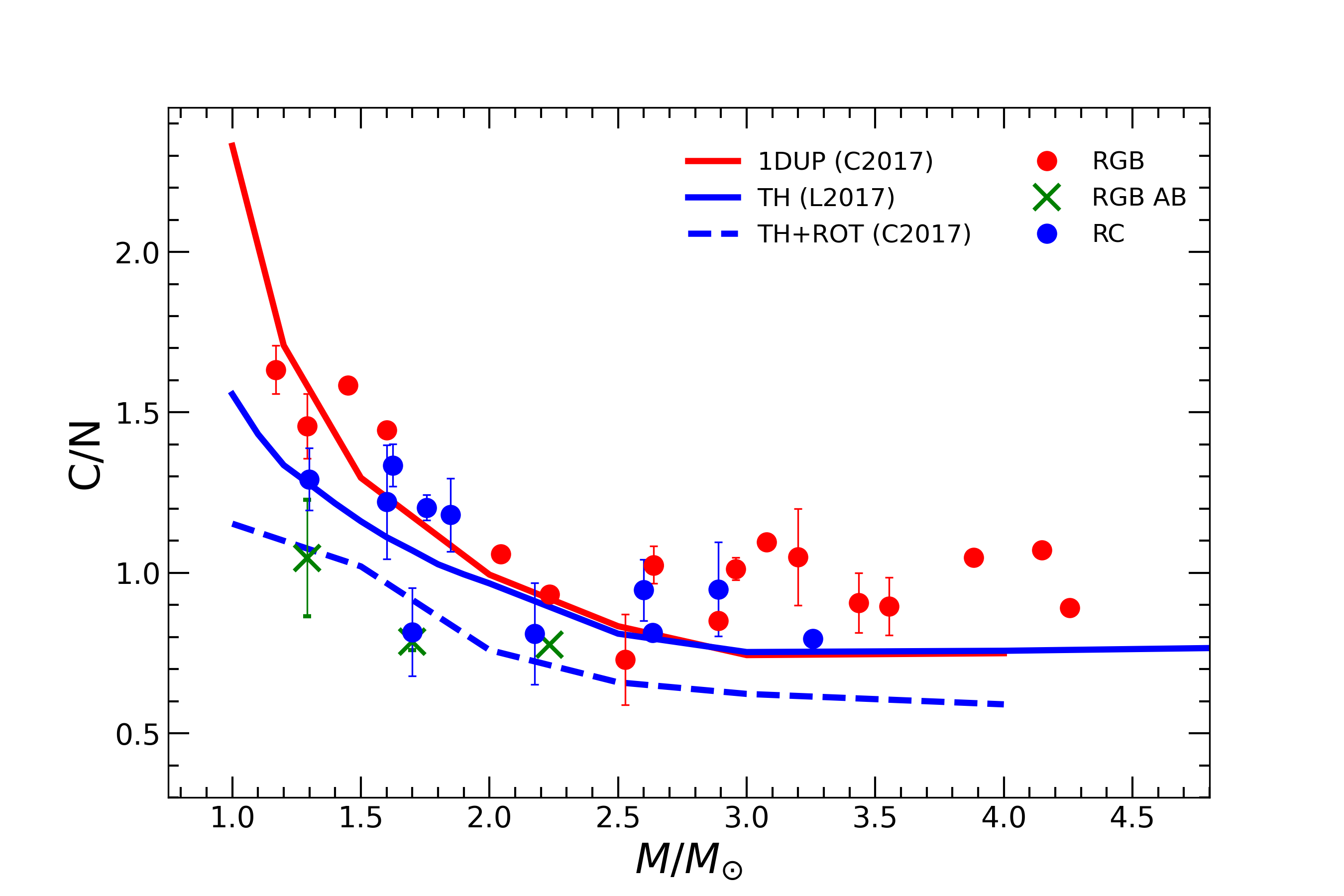}
    \caption{Comparison of mean C/N ratio values in open clusters for stars at different evolutionary stages with theoretical models. The red dots denote the mean values for RGB stars, the blue dots indicate the RC stars, and the green crosses represent the RGB stars above the luminosity bump (the scatter is presented when is available). The solid red line shows the predicted C/N ratios after 1DUP \citep{Charbonnel2017}. The solid blue line indicates the minimum C/N values predicted by models including thermohaline mixing \citep{2017A&A...601A..27L}. The dashed line corresponds to the model that includes both thermohaline- and rotation-induced mixing \citep{Charbonnel2017}.}
    \label{fig:CN_mass_errors}
\end{figure}

Recent stellar evolution models incorporating these hydrodynamical instabilities \citep{Lagarde12,2017A&A...601A..27L,Lagarde2019, Charbonnel2017}  have been designed to reproduce the observed abundance trends for C and N in evolved giants. In Fig.~\ref{fig:CN_mass_errors} we compare our determined C/N ratios as a function of turn-off mass with the theoretical models by \cite{Charbonnel2017} and \cite{2017A&A...601A..27L}. Two insights are gained from this comparison. 

By dividing our sample into distinct evolutionary phases of RGB stars below and above the RGB luminosity bump and RC stars, we attempted to effectively disentangle the effects of the standard 1DUP from subsequent non-standard mixing processes. 
This separation is crucial, as the dependence of the C/N ratio on turn-off mass is strictly modulated by a star's evolutionary state.

First, in the first-ascent giant stars, particularly those with larger turn-off masses, the observed C/N ratios are slightly higher than those predicted by standard models. We find that the average value of C/N (0.98~dex) determined in this work for the open cluster stars with turn-off masses larger than 2.8~$M_{\odot}$ is higher by about 0.20~dex compared to the models. This implies that these stars are less affected by the 1DUP than theoretically anticipated. A similar offset was noted by \cite{2025A&A...703A...4T} in their analysis of the $Gaia$-ESO survey data, suggesting that current models may slightly overestimate the mixing efficiency or mixing depth during the standard dredge-up for intermediate-mass stars.

Second, while the thermohaline instability is proposed to be the driving force behind extra mixing in the advanced stages of low-mass stars, we inferred that models relying on pure thermohaline mixing alone are insufficient to explain the full extent of the C/N depletion in RC and RGB AB stars of several open clusters in our sample. The rotation-induced or other extra-mixing mechanisms could be responsible for the relatively lower observed C/N ratios. 

\subsection{[C/N] as an age indicator}

The utility of the surface [C/N] abundance ratio as an age diagnostic -- termed a `chemical clock' -- stems from the strong mass dependence of the 1DUP. Since the depth of the convective envelope and the resulting surface pollution depend on stellar mass and the lifetime of a star is strictly determined by its mass, a correlation emerges between [C/N] and stellar age.

This relationship was empirically calibrated by \citet{Martig16} using a sample of field red giants with precise masses determined from \textit{Kepler} asteroseismology and abundances from the APOGEE survey (APOKASC sample; \citealt{pinsonneault14}). Subsequently, calibrations based on open clusters, which provide independent and precise ages, were developed. \citet{2019A&A...629A..62C} calibrated the relation using data from the \textit{Gaia}-ESO Survey \citep{gilmore22, randich22} and APOGEE DR14 \citep{majewski17, blanton17}, while \citet{2022AJ....163..229S} extended this work using the larger APOGEE DR17 dataset \citep{abdurrouf22}.

These studies differ in their treatment of different evolutionary phases. As discussed in Sect.~\ref{model_comparison}, extra mixing events (e.g. thermohaline and rotational mixing) further alter carbon and nitrogen  abundances after the RGB luminosity bump. Consequently, RC stars are expected to display lower [C/N] ratios than RGB stars of the same age and metallicity. This effect can be seen clearly in the [C/N] relations obtained for 44 open clusters in the $Gaia$-ESO survey \citep{2025A&A...703A...4T}. 

\begin{figure}
    \centering
    \includegraphics[width=1.0\columnwidth]{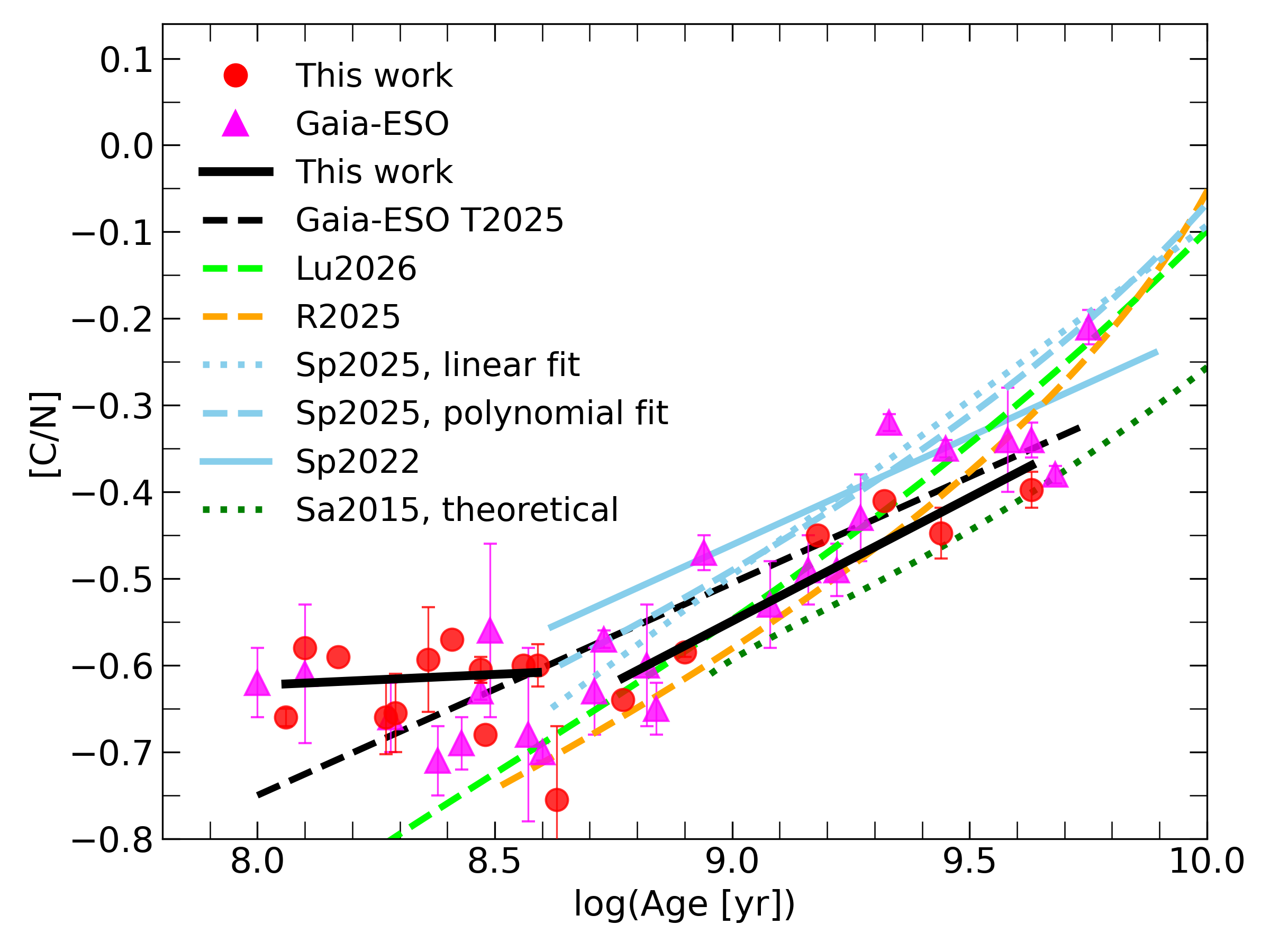}
    \caption{[C/N] versus age for the lower RGB stars. The dots represent the averaged [C/N] values and the scatter per cluster of this work and the obtained relation with the age (black line). The clusters shown without scatter bars contain only a single investigated star. For a comparison, the corresponding data and relation obtained in the $Gaia$-ESO survey, determined using the same method of analysis, are shown by the triangles and dashed black line \citep{2025A&A...703A...4T}. The relations obtained in the APOGEE survey for the open and globular clusters (\citealt{Spoo2025}) and for the field stars (\citealt{Lu2026, Roberts25}) are plotted as well. The dashed green line represents the theoretical relation by \cite{salaris2015} .  
    }
    \label{fig:CN_logage otherr_RGB}
\end{figure}

\begin{figure}
    \centering
    \includegraphics[width=1.0\columnwidth]{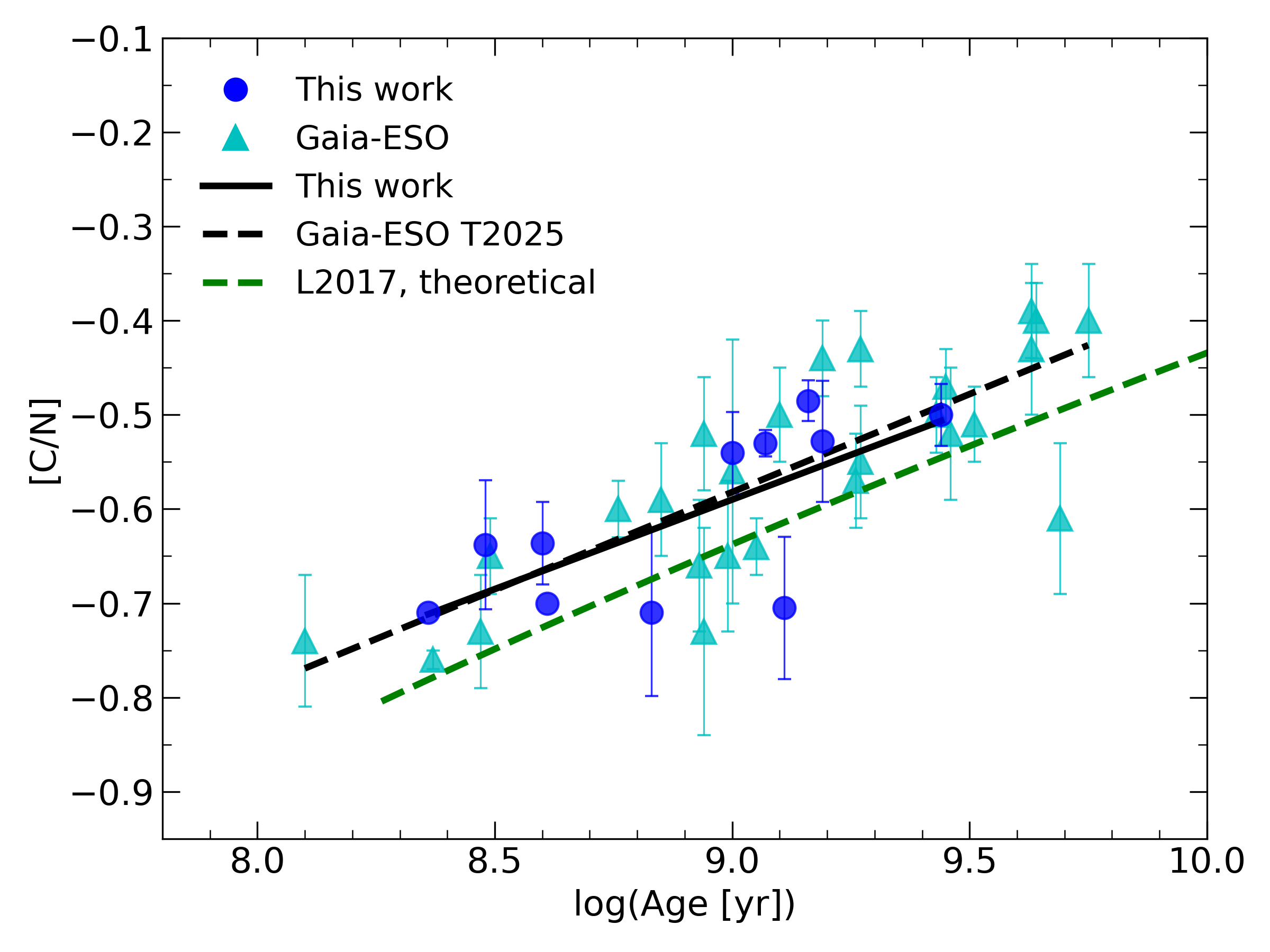}
    \caption{[C/N] versus age for the RC stars. The dots represent the averaged [C/N] values and the scatter per cluster of this work and the obtained relation with the age (black continuous line). The clusters shown without scatter bars contain only a single investigated star. For a comparison, the corresponding data and relation obtained in the $Gaia$-ESO survey, determined using the same method of analysis, are shown by the triangles and dashed black line \citep{2025A&A...703A...4T}. The dashed green line represents the theoretical relation by \cite{2017A&A...601A..27L}. 
    }
    \label{fig:CN_logage_RC}
\end{figure}

To evaluate the [C/N] ratio as an age indicator, we calculated individual linear regressions for the distinct evolutionary populations identified in our sample. For RGB stars, the clusters were separated into two groups at the boundary of 8.7~Gyrs as well. The resulting fits are illustrated in Figs.~\ref{fig:CN_logage otherr_RGB} and \ref{fig:CN_logage_RC}. The [C/N] ratio in younger clusters does not exhibit correlation with age, which is expected from stellar evolutionary models, as the C/N versus stellar mass relation becomes flat for stars with masses larger than $\sim3 M_\odot$. The resulting fits for RGB stars in older open clusters and RC stars are as follows: 

\begin{itemize}

    \item
For the RGB stars below the luminosity bump in older open clusters, \\
log\,Age = 3.509 $\cdot$ [C/N] + 10.926, PCC=0.93. 
\\
\item
For the core He-burning stars,\\
log\,Age = 5.263 $\cdot$ [C/N] + 12.105, PCC=0.71.
\end{itemize}

For convenience, we present the averaged [C/N] and C/N ratios and a number of stars in the lower-RGB and RC evolutionary stages for each open cluster (Table~\ref{tab:averagedCN}) used for calculations and for plotting data in the figures. NGC\,7209 was not included in the calculation due to its discrepantly low [C/N] ratio of unknown origin.

\begin{table*}
\centering
\caption{Averaged [C/N] and C/N ratios and a number of stars in the lower-RGB and RC evolutionary stages. }
\label{tab:averagedCN}
\begin{tabular}{lllcllc}
\hline\hline
Cluster name & [C/N] & C/N & N & [C/N] & C/N & N  \\
             &  RGB\,BB & RGB\,BB &  & RC & RC &  \\ 
\hline
Alessi\,1       &                &                &   & $-0.49\pm 0.02$ & $1.34\pm 0.07$ & 4 \\
Basel\,11b      & $-0.59\pm 0.06$ & $1.05\pm 0.15$ & 3 &                 &                &   \\
COIN-Gaia\,30   & $-0.57 $        & 1.10           & 1 &                 &                &   \\
Collinder\,350  &  $-0.64 $        & 0.93           & 1 &                 &                &   \\
Collinder\,463  & $-0.66\pm 0.01$ & $0.89\pm 0.02$ & 2 &                 &                &   \\
Gulliver\,51   &  $-0.60 $        & 1.02           & 1 &                 &                &   \\
IC\,4756     &                  &                  &   & $-0.71\pm 0.08$ & $0.81\pm 0.14$ & 10 \\
NGC\,2437    &  $-0.68$          & $0.85$            & 1 & $-0.64\pm 0.07$ & $0.95\pm 0.15$ & 5 \\
NGC\,2509    &   $-0.45 $        & 1.44           & 1 &                 &                &   \\
NGC\,2548    &   $-0.60\pm 0.02$ & $1.02\pm 0.06$ & 3 &                 &                &   \\
NGC\,2632    &                  &                &   & $-0.71\pm 0.09$ & $0.81\pm 0.16$ & 4 \\
NGC\,2682    &  $-0.40\pm 0.02$ & $1.63\pm 0.08$ & 4 &                 &                &   \\  
NGC\,6800    &                 &                &   & $-0.70$          & $0.81 $         & 1 \\
NGC\,6991    &                 &                &   & $-0.53\pm 0.06$ & $1.22\pm 0.18$ & 5 \\
NGC\,7086    &  $-0.66\pm 0.05$ & $0.91\pm 0.09$ & 2 &                 &                &   \\
NGC\,7209    &  $-0.76\pm 0.09$ & $0.73\pm 0.14$ & 2 &                 &                &   \\
NGC\,752      &                  &                  &   & $-0.53\pm 0.02$ & $1.20\pm 0.04$ & 3 \\
Ruprecht\,171 & $-0.45\pm 0.03$ & $1.46\pm 0.10$ & 4 &                 &                &   \\
Stock\,2       &                &                &   & $-0.64\pm 0.04$ & $0.94\pm 0.09$ & 8 \\
Tombaugh\,5    & $-0.66\pm 0.04$ & $0.90\pm 0.09$ & 3 &                 &                &   \\ 
UBC\,3         &   $-0.58 $        & 1.07           & 1 &                 &                &   \\ 
UBC\,60        &   $-0.59\pm 0.01$ & $1.06\pm 0.01$ & 2 &                 &                &   \\ 
UBC\,131       &                 &                &   & $-0.54\pm 0.04$ & $1.18\pm 0.11$ & 3 \\
UBC\,141       &  $-0.41 $        & 1.58           & 1 &                 &                &   \\ 
UBC\,169       &  $-0.61\pm 0.02$ & $1.01\pm 0.03$ & 2 &                 &                &   \\
UBC\,194       &               &                &   & $-0.71$          & 0.79           & 1 \\
UBC\,577       &                &                &   & $-0.50\pm 0.03$ & $1.29\pm 0.09$ & 3 \\     
UPK\,219       & $-0.59 $        & 1.05           & 1 &                 &                &   \\
\hline
\end{tabular}
\end{table*}

We corroborated the separate [C/N] relations with age for RGB and RC stars and the trends obtained in the $Gaia$-ESO survey of open clusters by \citet{2025A&A...703A...4T}, especially for RC stars. In Fig.~\ref{fig:CN_logage otherr_RGB}, we show the relations obtained in the recent study of RGB stars in open clusters by \cite{Spoo2025} and in studies of field RGB stars by \cite{Roberts25} and \cite{Lu2026}. \cite{Spoo2025} obtained a steeper trend, most probably due to the inclusion of old metal-deficient globular clusters in the open cluster sample. However, old low-metallicity stars have larger [C/N] values \citep[e.g.][]{salaris2015}, thus causing  the steepening of the relation. \cite{Roberts25} investigated field stars and computed relations that included [C/N] values reaching +0.15~dex; however, another study of the same APOGEE DR17 data by \cite{Lu2026} inferred that the [C/N] values above $-0.05$~dex could be dominated by merger products. It is not clear why the relation obtained for field stars by \cite{Lu2026} is quite steep as well. When only open clusters were used to determine the [C/N] versus age relation using the same APOGEE DR17 data by \cite{2022AJ....163..229S}, the relation was very similar to the relation obtained in our work and that in \cite{2025A&A...703A...4T}.  

As our study includes a large number of young open clusters with RGB stars investigated, it can be clearly seen for the younger first-ascent giant stars of larger turn-off masses that the [C/N] versus age relation becomes flatter, as expected from stellar evolutionary models. This behaviour challenges the quite steep relations obtained in the young age regime by the previous studies. 

Concerning the [C/N] relation with age in the oldest ages regime, open clusters are usually too young to be a reliable source of information. Here, data of field stars are necessary. The very recent study by \cite{Pakstiene2026} addresses this question in detail.

\subsection{Theia\,1214 association stars}

In this work, we have investigated two stars attributed to the Theia\,1214 association by  \cite{kounkel19}. As $Gaia$ does not have radial velocities for all of the stars in the Theia catalogue and the uncertainties for the existing measurements are large, it was difficult to determine which stars in this compact group are real members and which are spurious. A spectroscopic follow-up was proposed to distinguish between them on the basis of coherence in radial velocity and chemical composition. Having in mind that the open cluster NGC\,752 has signs of either dynamical or tidal dispersal \citep{Carraro2014} and features long, asymmetric tidal tails extending over 260~pc in the sky \citep{bhattacharya21, Boffin2022,Kos2024}, \cite{2025A&A...701A.289D} performed observations of two stars to test their possible relation to NGC\,752, and we determined the CNO abundances for these stars in this work. 

Although the two observed Theia stars fall well in the $G$ versus $G_{\rm BP}-G_{\rm RP}$ diagram to the location of the RC stars of NGC\,752, their CNO abundances and the atmospheric parameters differ essentially from each other and from the RC stars investigated in NGC\,752. To further investigate, we performed an independent age determination for all targets using UniDAM \citep{2018A&A...618A..54M}. For the primary NGC\,752 targets, we derived ages ranging from 0.74 to 1.20~Gyr, showing excellent agreement with the accepted age of 1.17~Gyr (log(Age[yr]) = 9.07) taken from \citealt{2020A&A...640A...1C}) and validating the reliability of our isochrone fitting. In contrast, the Theia\,1214 candidates revealed significant age discrepancies, both with respect to NGC\,752 and to each other. The derived ages are $\sim$0.5~Gyr for Theia\,1214\_1 and $>6$~Gyr for Theia\,1214\_3, a dispersion incompatible with a coeval stellar string. Given the combined evidence of discordant kinematics, chemical composition, and evolutionary ages, we classify the investigated Theia\,1214 targets as field stars that are not related to NGC\,752. Their membership in the same string is also questionable.

\section{Summary and conclusions }
\label{sec:summary_conclusions}
In this work, we have investigated CNO abundances in 88 stars of 28 open clusters and 2 stars of the association Theia\,1214 as a part of the SPA programme to calibrate the [C/N]--age relation for evolved stars. We separated our stars after the 1DUP episode into three subsamples: RGB stars below the luminosity bump, RGB stars above the luminosity bump (RGB AB) stars, and RC stars. We investigated the extent of mixing processes that occur during each phase and compared the observations with theoretical predictions. Using high-precision carbon, nitrogen, and oxygen abundances, we determined the individual [C/N]--age relations for the lower RBG and RC stars and compared them to theoretical models and empirically derived relations in other studies. Our results can be summarised as follows: 

\begin{itemize}
\item
The RC versus age relation is systematically offset towards lower [C/N] values compared to the RGB trend.  This provides direct observational evidence of the efficiency of extra mixing mechanisms (e.g. thermohaline- and rotation-induced mixing) that act during the post-RGB luminosity bump evolution. These results support the conclusion that a single chemical clock calibration cannot be applied indiscriminately to all giant stars. Precise age dating requires the separation of first-ascent giants from the RC stars to account for the additional carbon and nitrogen abundance alterations in the latter.

\item
In the first-ascent giant stars with larger turn-off masses, the observed C/N ratios are slightly higher than those predicted by standard models.

\item
For the younger first-ascent giant stars that are of larger turn-off masses, the [C/N] versus age relation becomes flatter as expected from stellar evolutionary models.

\item 
The two stars of the Theia\,1214 association investigated in this work, according to their chemical composition, age, and kinematics, can hardly be related to the open cluster NGC\,752 and be members of the same string. 
\end{itemize}

\section*{Data availability}

Table~\ref{tab:cno_abundances} is only available in electronic form at the CDS via anonymous ftp to cdsarc.u-strasbg.fr (130.79.128.5) or via \url{http://cdsweb.u-strasbg.fr/cgi-bin/qcat?J/A+A/}

\begin{acknowledgements}
This study is based on observations made with the Italian Telescopio Nazionale
Galileo (TNG) operated on the island of La Palma by the Fundación
Galileo Galilei of the INAF (Istituto Nazionale di Astrofisica) at the
Observatorio del Roque de los Muchachos. This research used facilities of the Italian Center for Astronomical Archive (IA2) operated by INAF at the Astronomical Observatory of Trieste. This work has made use of data from the European Space Agency (ESA) mission $Gaia$ (https://www.cosmos.esa.int/gaia), processed by the Gaia Data Processing and Analysis Consortium (DPAC, https://www.cosmos.esa.int/web/gaia/dpac/consortium). Funding for the DPAC has been provided by national institutions, in particular the institutions participating in the Gaia Multilateral Agreement. We thank the referee for suggestions that improved the presentation and clarity of the article.
B.Ć. and A.D. acknowledge funding from the Research Council of Lithuania (LMTLT, grant No. S-LL-24-4).
A.B, V.D'O, M.D.P. acknowledge funding from INAF MiniGrant 2022
(High resolution spectroscopy of open clusters) and the INAF grant Open Clusters and stellar structures in the local Galactic disk (ref. Antonella Vallenari) (CRA 1.05.23.05.19).

\end{acknowledgements}

\bibliographystyle{aa} 
\bibliography{biblio.bib} 

\end{document}